\documentclass[aps,pra,twocolumn,floatfix,reprint,showpacs,superscriptaddress]{revtex4-1}
\usepackage{amsmath}
\usepackage{color}
\usepackage{graphicx}
\usepackage{float}
\usepackage{braket}

\begin{document}

\title{Entangled microwaves as a resource for entangling spatially separate solid-state qubits: superconducting qubits, NV centers and magnetic molecules}

\author{Angela Viviana G\'omez}
\email{av.gomez176@uniandes.edu.co}
\affiliation{Departamento de F\'{\i}sica, Universidad de Los Andes, A.A.4976, Bogot\'a D.C., Colombia}
\affiliation{Instituto de F{\'\i}sica Fundamental, IFF-CSIC, Calle Serrano 113b, Madrid E-28006, Spain}

\author{Ferney Javier Rodr\'{\i}guez}
\affiliation{Departamento de F\'{\i}sica, Universidad de Los Andes, A.A.4976, Bogot\'a D.C., Colombia}

\author{Luis Quiroga}
\affiliation{Departamento de F\'{\i}sica, Universidad de Los Andes, A.A.4976, Bogot\'a D.C., Colombia}

\author{Juan Jos\'e Garc\'{\i}a-Ripoll}
\affiliation{Instituto de F{\'\i}sica Fundamental, IFF-CSIC, Calle Serrano 113b, Madrid E-28006, Spain}

\date{\today }

\begin{abstract}
Quantum correlations present in a broadband two-line squeezed microwave state can induce entanglement in a spatially separated bipartite system consisting of either two single qubits or two qubit ensembles. By using an appropriate master equation for a bipartite quantum system in contact with two separate but entangled baths, the generating entanglement process in spatially separated quantum systems is thoroughly characterized. Decoherence thermal effects on the entanglement transfer are also discussed. Our results provide evidence that this entanglement transfer by dissipation is feasible yielding to a steady-state amount of entanglement in the bipartite quantum system which can be optimized for a wide range of realistic physical systems that include state-of-the-art experiments with NV centers in diamond, superconducting qubits or even magnetic molecules embedded in a crystalline matrix.
\end{abstract}

\pacs{03.67.Mn,03.65.Ud,03.67.Lx}
\maketitle

\section{INTRODUCTION}
The generation and preservation of entanglement is one of the basic ingredients in many scalable quantum information protocols. Quantum cryptography~\cite{Ekert91,Gisin02}, quantum communication~\cite{Bennet93}, quantum repeaters and certain models of quantum computation~\cite{Briegel09,Raussendorf01,Gottesman99}, demand preexis-ting entangled states, either at short distances or at long separations. If we focus on the establishment of pairwise entanglement, there exist three basic approaches: (i) an interaction in some past moment~\cite{Turchette98}, (ii) a joint measurement with an entangled state as an outcome~\cite{Moehring07} or (iii) an interaction with a third party or mediator, such as phonons~\cite{SchmidtKaler03} or photons~\cite{Julsgaard01,Chou05,Matsukevich06,Schug13}, and which often can be reinterpreted as (ii) once the mediator is traced out.

We have cited some examples of Atomic and Molecular Physics experiments where all these ideas have been put into practice. However, in recent years the field of solid-state quantum information processing has reached a status in which many of those entanglement protocols can be competitively reproduced, with similar goals and rapidly improving performance, using semiconductor quantum dots~\cite{Gao12}, Nitrogen Vacancy (NV) centers in diamond~\cite{Neumann10,Bernien13,Hensen15}, superconducting qubits~\cite{Ansmann09,Chow10,Baur12}, surface plasmon polaritons~\cite{Pino14} or superconducting microwave photons~\cite{Eichler11,Menzel12,Flurin12,Lang13,Eichler14}, to name a few possibilities. In this context, a remarkable idea is the hybridization of different technologies in a single setup, thus synthesizing the best of each. One attractive example is the integration of superconducting resonators with NV centers in diamond. These systems exploit the long coherence times of the NV spin in diamond jointly with the promise of high scalabi-lity and robust control of SC circuits~\cite{Nori,Marcos,Zhu,Kubo}. Experimentally, the strong coupling between a spin ensemble and a superconducting resonator has been demonstrated in the linear or Gaussian regime~\cite{Amsuss,Awschalom,Wach}, where the resonator and spin ensemble are both modeled as interacting harmonic oscillators. In addition, the strong coupling in this hybrid systems has allowed to transfer the state between the NV ensemble and a superconducting resonator~\cite{Stevepra} while some other works show an improvement in the coherence times and the transfer of single excitations with a flux qubit~\cite{Munro}.

\begin{figure}[t]
\includegraphics[width=0.55\textwidth]{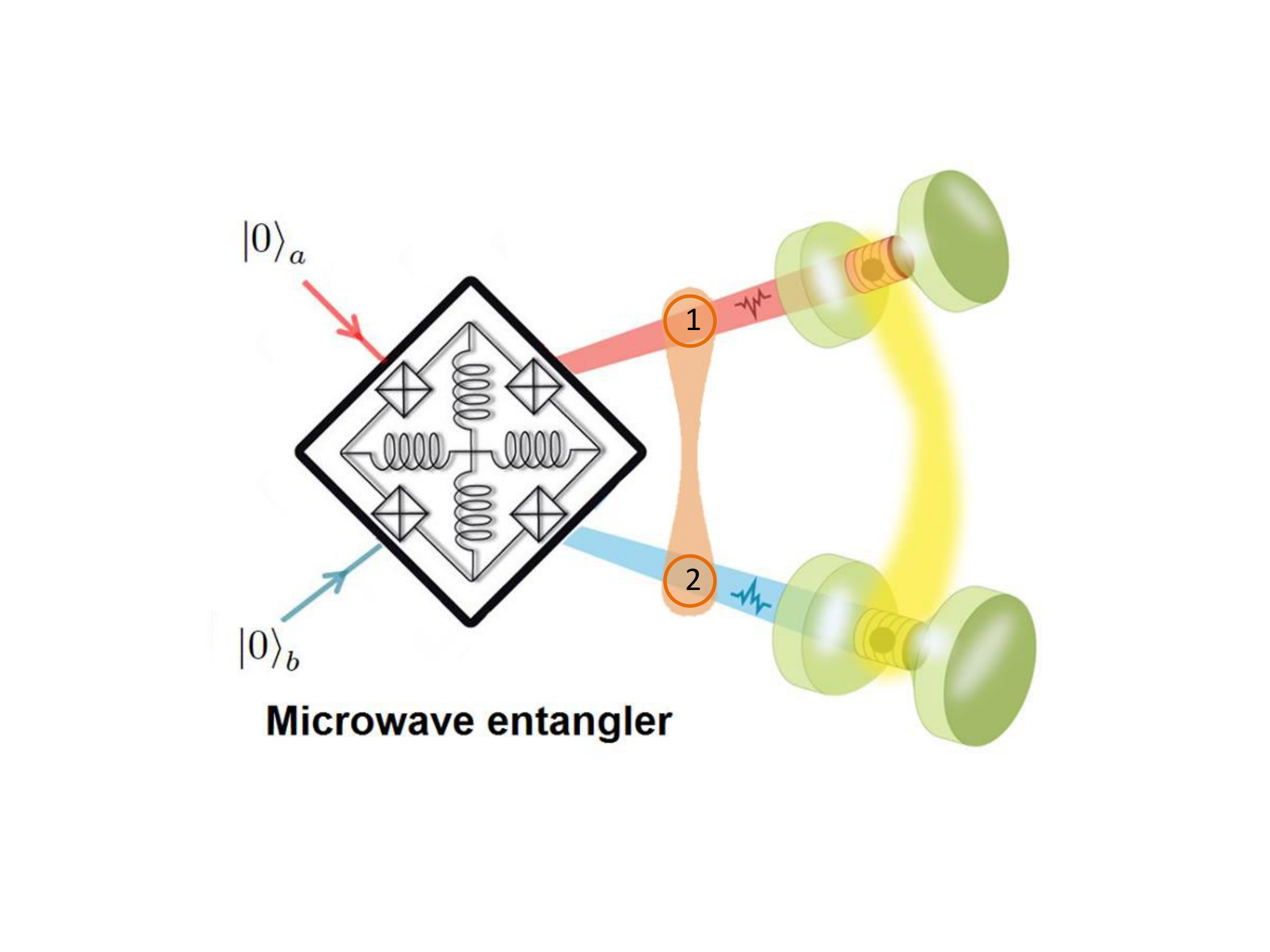}
\caption{\textbf{System:} Two spatially separated qubits (or two qubit ensembles) in different branches coupled with entangled microwaves generated by a parametric Josephson amplifier.}
\label{fig:setup}
\end{figure}

A wealth of studies
have clarified the transfer of entanglement from infinite-dimensional field systems to discrete matter systems, especially those involving driven cavities with embedded qubits \cite{Solano,Cirac}. A first proposal was limited to consider the unitary evolution of separate microcavity plus single qubit, with the radiation fields in a highly squeezed pure state ~\cite{Paternostro04}. Shortly afterwards, a highlighted scenario where two remote single-mode cavities containing a single qubit each was proposed to reach maximally entangled two-qubit states in both transient and steady-state regimes ~\cite{Adesso2010}, by driving the cavities with highly entangled broadband two-mode Gaussian fields which act as local environments for each qubit. A standard formalism of second-order perturbation theory (Born-Markov approximation) allows to determine the sufficient and necessary conditions to  reach a successful entanglement transfer from a highly mixed but entangled broadband multi-mode reservoir to a spatially separate qubit pair ~\cite{Paternostro2004}.
Finally, pure and mixed entangled fields have also been proposed to quantum correlate pairs of other initially uncorrelated subsystems \cite{Zippilli2014} especially when a mechanism for the replication over many matter subsystem pairs can be identified ~\cite{Zippilli2013}. All of these protocols for the controlled manipulation of the entanglement distribution represent important steps towards the engineering of quantum networks. Motivated by this joint progress from both the theoretical and experimental advances, in this work we study the transfer of entanglement from a continuous, broadband,
two-line squeezed microwave field (TLSMF) as generated by Josephson Parametric Amplifiers (JPA)~\cite{Yurke87},
onto a bipartite system consisting of two qubits or two spin ensembles ---either which can be made of NV-centers, mole-cular magnets~\cite{Juanjo} or superconducting qubits---, roughly as sketched in Fig.~\ref{fig:setup}. By contrast with previously des-cribed protocols here we propose a new scheme for gene-rating entanglement between spatially remote qubit systems exploiting a setup which accommodates the peculiarities of circuit-QED and of the novel field of propagating quantum microwaves, without resorting to embed the qubits in a microcavity but leaving them to couple directly with the TLSMF which is being continuously replenished. Moreover, we allow the qubit subsystems to be in contact with local environments which provide fully incoherent processes (thermal decoherence) which compete with the quantum coherent generating processes as represented by the TLSMF. In this way, we provide quantitative evidence for limitations on the entanglement transfer as caused by ubiquitous thermal events, a shortcoming of previous studies ~\cite{Adesso2010,Paternostro2004}.
We thus study the pure TLSMF entanglement transfer power as well as its limitations set by additional couplings to thermal baths, extending earlier results for quantum systems in contact with several bath fields~\cite{quiroga07,castillo13}, using Markovian master equations.

The paper is organized as follows: Sect.\ \ref{sec:theory} describes the general formalism for addressing the entanglement driving by two dissipative entangled baths as represented by TLSMF. The case of a matter subsystem corresponding to two single separate qubits is developed in Sect.\ \ref{sec:qubits} while Sect.\ \ref{sec:ensembles} is devoted to two distant qubit ensembles. The solutions discussed in those sections encompass both resonant and non-resonant cases between central microwave frequency and matter. Further decoherence effects on the qubit pair or qubit ensembles are considered in Sect.\ \ref{sec:decoherence}. Realistic solid-state implementations are explored in Sect.\ \ref{sec:implementation} where experimental parameters as appropriate to NV centers in diamond, superconducting qubits and magnetic molecules are considered. Finally, Sect.\ \ref{sec:summary} summarizes our main conclusions, while technical details are relegated to the Appendix.

\section{BIPARTITE QUANTUM SYSTEM IN CONTACT WITH TWO SEPARATE BATHS}\label{sec:theory}
We start by putting on theoretical grounds the formalism yielding to the master equation describing the generation of quantum correlations on a quantum matter bipartite system by the driving from two separate entangled (microwave) reservoirs. In this section we limit ourselves to the effects of the TLSMF on the matter subsystem. Other couplings of the matter with additional reservoirs providing extra matter decoherence channels are discussed below, see Sect.\ \ref{sec:decoherence}. Here we follow and extend to squeezed reservoirs previous results from~\cite{quiroga07,castillo13} which have been already applied to a quantum system coupled to two thermal reservoirs at different temperatures. We consider a composed quantum system, including the baths, formed by two spatially separated lines or branches as depicted in Fig.~\ref{fig:setup}.
In each branch a part of a multi-squeezed microwave field interacts locally with a quantum system of interest. Thus, the full Hamiltonian reads as:

\begin{equation}
\hat{H}=\sum_{j=1}^2\hat{H}_j=\sum_{j=1}^2\left ( \hat{Q}_j+\hat{R}_{j}+\hat{V}_{j}\right ) \label{Eq:eq1}
\end{equation}
where for arm $j$, $\hat{Q}_j$, $\hat{R}_{j}$ and $\hat{V}_{j}$ denote the partial Hamiltonians for the quantum or matter system itself, the free microwave radiation field and the matter-radiation interaction terms, respectively. Notice that $\left [ \hat{H}_1,\hat{H}_2 \right ]=0$.

The aim is to find the equation of motion for the quantum system reduced density operator, $\hat{\rho}(t)$, from the unitary evolution of the full super-system density operator $\hat{\gamma}(t)$. To proceed further we express the full dynamics in the interaction picture given by the transformation
\begin{equation}
\hat{U}^{\dag }(t)=\hat{U}_1^{\dag }(t)\hat{U}_2^{\dag }(t)=\prod_{j=1}^2e^{i\left(\hat{Q}_j+\hat{R}_{j}\right) t} \label{Eq:eq3}
\end{equation}
such that an interaction picture operator $\hat{O}_I(t)$ is connected with its Schr\"{o}dinger version $\hat{O}_S$ by
$\hat{O}_I(t)=\hat{U}^{\dag }(t)\hat{O}_S\hat{U}(t)$.
The full super-system (bipartite quantum system + reservoirs) density
operator satisfy the Liouville-Von Neumann equation ($\hbar=1$)
\begin{equation}
\frac{d\hat{\gamma}_I(t)}{dt}=-i[\hat{V}_I(t),\hat{\gamma}_I(t)] \label{Eq:Liu}
\end{equation}
with $\hat{V}_I(t)=\hat{V}_{1,I}(t)+\hat{V}_{2,I}(t)$.
We assume that the coupling strength between the central quantum matter system and the microwave reservoirs is weak enough to express $\hat{\gamma}(t)$ as
\begin{equation}
\hat{\gamma}_I(t)=\hat{\rho}_I(t)\otimes \hat{\rho}_{1,2}^{B} \label{Eq:eq2}
\end{equation}
where the baths are described by a stationary correlated (non-separable) density matrix $\hat{\rho}_{1,2}^{B}$.
Thus, up to second order in the matter-radiation interaction strength, we obtain~\cite{quiroga07,castillo13}:
\begin{equation}
\frac{d\hat{\rho}_I(t)}{dt}=(-i)^{2}\int_{0}^{t}dt_{1}Tr_{R}\left \{[\hat{V}_I(t),[\hat{V}_I(t_{1}),\hat{\rho}_I(t_1)\otimes\hat{\rho}_{1,2}^{B}
]]\right \} \label{Eq:eq10}
\end{equation}
where $Tr_R\left \{ ... \right \}$ denotes the partial trace over the squeezed microwave radiation reservoirs.

According to ~\cite{Flurin12} the broadband TLSMF produced by a JPA (see Fig.~\ref{fig:setup}) can be described by $\left\vert S_{q}\right\rangle =\hat{S}\left\vert \{0\}_{1}\right\rangle
\otimes \left\vert \{0\}_{2}\right\rangle $ ~\cite{scully},

where the two arms multi-mode vacuum state is denoted as $|\{ 0\}_1\rangle\otimes|\{ 0\}_2\rangle$, in such a way that the stationary entangled baths are described by a non-separable density operator of the form
\begin{eqnarray}
\hat{\rho}_{1,2}^{B}=\hat{S}|\{ 0\}_1\rangle\otimes|\{ 0\}_2\rangle\langle\{ 0\}_2|\otimes\langle\{ 0\}_1|\hat{S}^{\dag}
\label{Eq:eq9}
\end{eqnarray}
indicating that in the arm $j$ a broadband multi-mode distribution centered on frequency $\omega_{Lj}$ is found.
The multi-mode squeezing operator is given by
\begin{widetext}
\begin{eqnarray}
\hat{S}=\mathrm{exp}\left \{\sum_{n,m} s(\omega_n,\omega_m) \left [ \hat{a}^{\dag}_{1}(\omega_{L1}+\omega_n)\hat{a}^{\dag}_{2}(\omega_{L2}-\omega_m)-
\hat{a}_{1}(\omega_{L1}+\omega_n)\hat{a}_{2}(\omega_{L2}-\omega_m) \right ]\right \} \label{Eq:eq8}
\end{eqnarray}
\end{widetext}
where $\hat{a}_{1}(\omega_{L1}+ \omega_n)$ and $\hat{a}_{2}(\omega_{L2}- \omega_n)$ denote the photon annihilation operators for mode $n$ of arms $j=1,2$, respectively. Although microwaves over a broadband continuum of modes are assumed, the mode indexes $n,m$ in Eq.~\eqref{Eq:eq8} are represented by discrete labels for simplicity. In Eq.~\eqref{Eq:eq8}, $s(\omega_n,\omega_m)=s_{n,m}$ are associated with the function (taken as real) of squeezing parameters between mode $\omega_{L1}+\omega_n$ in path $1$ and mode $\omega_{L2}-\omega_m$ in path $2$. The multi-mode entangled state given by Eq.~\eqref{Eq:eq9} describes two spatially separated but highly {\it entangled baths} that we will use as a resource for entangling the matter subsystems themselves.

The microwave reservoirs are described by local Hamiltonians in each arm such as
\begin{equation}
\hat{R}=\sum_{j=1}^2\hat{R}_{j}=\sum_{j=1}^2\sum_{n}\left ( \omega _{Lj}-(-1)^j\omega _{n}\right )\hat{a}_{n,j}^{\dag }\hat{a}_{n,j} \label{Eq:eq5}
\end{equation}
In the last equation we have explicitly written $\hat{a}_{n,1}$ and $\hat{a}_{n,2}$ in place of $\hat{a}_{1}(\omega_{L1}+ \omega_n)$ and $\hat{a}_{2}(\omega_{L2}- \omega_n)$, respectively, a double notation that we take as equivalent in the following. Additionally, the matter Hamiltonian can be written as
\begin{equation}
\hat{Q}=\sum_{j=1}^2\hat{Q}_{j}=\sum_{j=1}^2\omega_j \hat{q}_{j}^{+}\hat{q}_{j}^{-} \label{Eq:eq7}
\end{equation}
where $\hat{q}_{j}^{\pm}$ denote single excitation operators for the matter sub-system in branch $j$ and the commutation relation $\left [ \hat{q}_{j}^{+}\hat{q}_{j}^{-},\hat{q}_{j^{\prime}}^{\pm} \right ]=\pm \hat{q}_{j}^{\pm}\delta_{j,j^{\prime}}$ should hold. The specific physical meaning of $\omega_j$ and $\hat{q}_{j}^{\pm}$ in Eq.~\eqref{Eq:eq7} will be discussed below for different cases. Finally, the matter-radiation interaction term in arm $j$ becomes
\begin{equation}
\hat{V}=\sum_{j=1}^2\hat{V}_{j}=\sum_{j=1}^2\sum_{n}g_{n,j}\left( \hat{q}_{j}^{+}\hat{a}_{n,j}+\hat{q}_{j}^{-}\hat{a}_{n,j}^{\dag }\right) \label{Eq:eq6}
\end{equation}
where $g_{n,j}=g_j(\omega_{Lj}-(-1)^j\omega_n)$ is the coupling strength between matter sub-system and mode $n$ in branch $j$.
Therefore,
the interaction picture expression for the matter-radiation coupling Hamiltonian takes the form
\begin{widetext}
\begin{eqnarray}
\nonumber \hat{V}_I(t)&=&\sum_{n}g_1(\omega_{L1}+\omega_n)\left( \hat{q}_{1}^{+}\hat{a}_{n,1}e^{i\left ( \omega_1-\omega_{L1}-\omega_n \right )t}+\hat{q}_{1}^{-}\hat{a}_{n,1}^{\dag }e^{-i\left ( \omega_1-\omega_{L1}-\omega_n \right )t}\right)\\
&+& \sum_{n}g_2(\omega_{L2}-\omega_n)\left( \hat{q}_{2}^{+}\hat{a}_{n,2}e^{i\left ( \omega_2-\omega_{L2}+\omega_n \right )t}+\hat{q}_{2}^{-}\hat{a}_{n,2}^{\dag }e^{-i\left ( \omega_2-\omega_{L2}+\omega_n \right )t}\right) \label{Eq:eq11}
\end{eqnarray}
\end{widetext}
By inserting Eq.~\eqref{Eq:eq11} into Eq.~\eqref{Eq:eq10},
expressions involving bath operators such as $Tr_{R_{1},R_{2}}\{\hat{\rho}_{1,2}^{B}\hat{a}
_{n,k}^{\pm }\hat{a}_{m,j}^{\pm }\}=\langle \hat{a}^{\pm}_{n,k}\,\hat{a}^{\pm}_{m,j}\rangle$ ($j,k=1,2$) should be evaluated, which for the entangled
bath density operator given by Eq.~\eqref{Eq:eq9} leads to
\begin{eqnarray}
\nonumber \langle \hat{a}^{\pm}_{n,k}\,\hat{a}^{\pm}_{m,j}\rangle=\langle \{ 0\}_1|\otimes\langle \{ 0\}_2|\,\hat{S}^{\dag}\,\hat{a}^{\pm}_{n,k}\,\hat{a}^{\pm}_{m,j}\,\hat{S}\,|\{ 0\}_1\rangle\otimes|\{ 0\}_2\rangle\\
\label{Eq:eq13}
\end{eqnarray}

\begin{figure}[t]
\begin{center}
\includegraphics[width=0.5\textwidth]{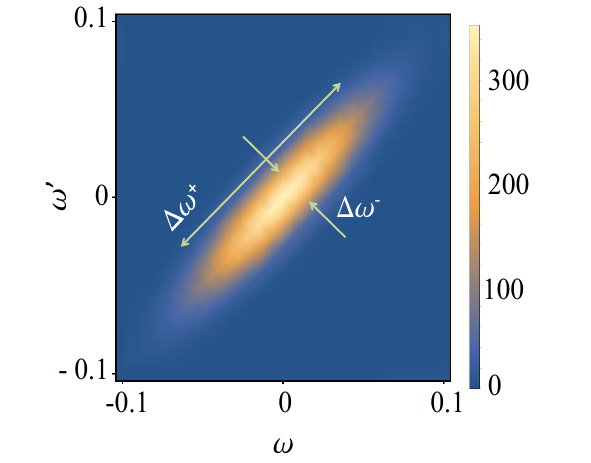}
\end{center}
\caption{Microwave squeezing strength $s(\omega,\omega')$ described by a double Gaussian function of two arm frequencies $\omega$ and $\omega'$ (in $\omega_{L}$ units, $\omega_{L1}=\omega_{L2}=\omega_{L}$) with $\Delta\omega^+=0.09\omega_{L}$ and $\Delta\omega^-=0.02\omega_{L}$.}
\label{fig:micro}
\end{figure}

The squeezing function $s_{n,m}$ in Eq.~\eqref{Eq:eq8} is assumed to be Gaussian \cite{Salazar12}, i.e.
\begin{eqnarray}
\nonumber s_{n,m}=\tilde{s} \left (\frac{2}{\pi \Delta\omega^- \Delta\omega^+}\right ) e^{-\left ( \frac{\omega_n-\omega_m}{\Delta\omega^{-}}\right )^2} e^{-\left ( \frac{\omega_n+\omega_m}{\Delta\omega^{+}}\right )^2}\\ \label{Eq:eq115}
\end{eqnarray}
where $\Delta\omega^-$ is associated with the width of the two-bath correlations as determined by the JPA pump duration while $\Delta\omega^+$ corresponds to the spectral width of the two arms coherence (see Fig.~\ref{fig:micro}). A special situation occurs when the two-bath correlations are perfect, i.e.
$\Delta\omega^{-}\rightarrow 0$, transforming the first Gaussian in Eq.~\eqref{Eq:eq115} in a delta function yielding to $s_{n,m}=\delta_{n,m}s_n$ with
\begin{eqnarray}
s_n=s_0 e^{-\left ( \frac{2\omega_n}{\Delta\omega^{+}} \right )^2} \label{Eq:eq116}
\end{eqnarray}
being $s_0$ the maximum squeezing between microwaves at central frequencies $\omega_{L1}$ and $\omega_{L2}$.
The reservoir correlations are then given by the expressions, see Appendix:
\begin{eqnarray}
\langle \hat{a}_{n,1}^{\dag }\hat{a}_{m,1}\rangle  &=&\langle \hat{a}_{n,2}^{\dag }\hat{a}_{m,2}\rangle=\delta_{n,m}\mathrm{sinh}^2(r_{n})  \notag \\
\langle \hat{a}_{n,1}^{\dag }\hat{a}_{m,2}^{\dag }\rangle  &=&\langle \hat{a}
_{n,1}\hat{a}_{m,2}\rangle =\delta_{n,m}\mathrm{sinh}(r_{n})\mathrm{cosh}(r_{n})  \notag \\
\langle \hat{a}_{n,1}\hat{a}_{m,1}^{\dag }\rangle  &=&\langle \hat{a}_{n,2}
\hat{a}_{m,2}^{\dag }\rangle =\delta_{n,m}\mathrm{cosh}^2(r_{n})  \label{Eq:eq118}
\end{eqnarray}

In the following, we shall assume that the frequency detunings ($\Delta_j=\omega_j-\omega_{Lj}$) between quantum matter ($\omega_j$) and central microwave distribution ($\omega_{Lj}$) satisfy the condition $\Delta_1=-\Delta_2=\Delta$, although non specific relation between $\omega_{1}$ and $\omega_{2}$ is required. Arbitrary detuning effects will be 
discussed elsewhere. Within the Wigner-Weisskopf approach, Eq.~\eqref{Eq:eq10} yields to a master equation in Lindblad form as:
\begin{equation}
\frac{d\hat{\rho}}{dt}=\hat{\mathcal{L}}_{MW}\hat{\rho}(t)
=\left [ \hat{\mathcal{L}}_{1}+\hat{\mathcal{L}}_{2}+\hat{\mathcal{L}}_{1,2}\right ]\hat{\rho}(t) \label{Eq:n1}
\end{equation}
where the MW label in the Lindbladian $\hat{\mathcal{L}}_{MW}$ reinforces the idea of just taking into account TLSMF effects for the moment.
The {\it local} TLSMF terms are given by ($j=1,2$)
\begin{widetext}\begin{eqnarray}
\nonumber \hat{\mathcal{L}}_j\hat{\rho}(t)&=&\Gamma_{j}(\Delta)\left \{ {\rm cosh}^2[s(\Delta)]\left [ 2\hat{q}_j^-\hat{\rho}(t)\hat{q}_j^+-\hat{q}_j^+\hat{q}_j^-\hat{\rho}(t)-\hat{\rho}(t)\hat{q}_j^+\hat{q}_j^- \right ] +
{\rm sinh}^2[s(\Delta)])\left [2\hat{q}_j^+\hat{\rho}(t)\hat{q}_j^- -\hat{q}_j^-\hat{q}_j^+\hat{\rho}(t)-\hat{\rho}(t)\hat{q}_j^-\hat{q}_j^+ \right ] \right \}\\
\label{Eq:nnn2}
\end{eqnarray}\end{widetext}
where, according to Eq.~\eqref{Eq:eq116} $s(\Delta)=s_0 e^{-\left ( \frac{2\Delta}{\Delta\omega^{+}} \right )^2}$,
and the effective {\it local} matter-radiation coupling becomes given by
\begin{eqnarray}
\Gamma_j(\Delta)=\sum_n \left |g_{j}(\omega_{Lj}-(-1)^j\omega_n)\right |^2\delta (\omega_n-\Delta)
\label{Eq:nv22}
\end{eqnarray}
Furthermore, the cross or {\it non-local} TLSMF Lindblad term in Eq.~\eqref{Eq:n1} reads as
\begin{widetext}\begin{eqnarray}
\nonumber \hat{\mathcal{L}}_{1,2}\hat{\rho}(t)&=&\Gamma_{1,2}(\Delta){\rm sinh}[2s(\Delta)]\left \{ \left [ 2\hat{q}_2^+ \hat{\rho}(t)\hat{q}_1^+-\hat{\rho}(t)\hat{q}_1^+\hat{q}_2^+-\hat{q}_1^+\hat{q}_2^+ \hat{\rho}(t)\right ]+ \left [2\hat{q}_1^+ \hat{\rho}(t)\hat{q}_2^+ -\hat{\rho}(t)\hat{q}_2^+\hat{q}_1^+-\hat{q}_2^+\hat{q}_1^+ \hat{\rho}(t)\right ]+H.C.\right \}\\
\label{Eq:n2}
\end{eqnarray}\end{widetext}
where the effective {\it non-local} matter-bath coupling constant in Eq.~\eqref{Eq:n2} is given by:
\begin{eqnarray}
\nonumber \Gamma_{1,2}(\Delta)=\sum_n g_{1}(\omega_{L1}+\omega_n)g_{2}(\omega_{L2}-\omega_n)\delta (\omega_n-\Delta)\\
\label{Eq:nv222}
\end{eqnarray}
and $H.C.$ denotes the hermitian conjugate terms. Lindblad terms such as $\hat{\mathcal{L}}_{j}$ in Eq.~\eqref{Eq:nnn2} denote the dissipative coupling of microwaves in line $j$ with its respective matter sub-system. These local dissipative terms have similar forms as the coupling of a single matter piece with a single thermal bath thus producing a null result for entanglement. Indeed, the most interesting dissipative term is the non-local Lindblad term $\hat{\mathcal{L}}_{1,2}$ in Eq.~\eqref{Eq:n2} which is the responsible for entangling the two matter sub-systems. In the next two sections we will apply this formalism to different physical systems, specifically those formed by two individual separate qubits as well as to different ensembles of qubits interacting with entangled microwave photons.

\section{BIPARTITE QUANTUM SYSTEM: A SOLID-STATE QUBIT PAIR}\label{sec:qubits}

The first setup under consideration is a hybrid combination of two separate single qubits, interacting with a broadband TLSMF. The solid-state qubits could be individual NV centers, magnetic nanomolecules or superconducting qubits, while the entangled microwave fields can be generated in a variety of ways from JPA devices. The qubits are represented by Pauli operators $\hat{\sigma}_{j,z}$ and $\hat{\sigma}_{j}^{\pm}$, with splitting energies $\omega _{j}$ ($j=1,2$). The qubit-radiation interaction strength is given by
$g_{n,j}=g_j(\omega_{Lj}-(-1)^j\omega_n)$ for the qubit in the transmission line $j$ coupled to mode $n$.
In order to quantify the entangling power of the microwave entangled reservoirs acting on the non-interacting qubit pair we start by writing down solutions to the master equation given in Eq.~\eqref{Eq:n1} with Lindblad terms as in Eq.~\eqref{Eq:nnn2} and Eq.~\eqref{Eq:n2} with the substitutions $\hat{q}_j^+ \rightarrow \hat{\sigma}_j^+$ and $\hat{q}_j^- \rightarrow \hat{\sigma}_j^-$. The effect of the squeezing between the baths on the qubit pair evolution is evident in the crossed Lindblad term $\hat{\mathcal{L}}_{1,2}$ where the presence of simultaneous two qubit excitation operators $\hat{\sigma}_{1}^{+}$ and $\hat{\sigma}_{2}^{+}$ (or their hermitian conjugates) occur. In a two-qubit base ordered as $\{|+,+\rangle ,|+,-\rangle ,|-,+\rangle ,|-,-\rangle \}$
the two-qubit density operator $\hat{\rho}(t)$ has the form:
\begin{equation}
\hat{\rho}(t)=\left(
\begin{array}{cccc}
\rho_{1,1}(t) & 0 & 0 & \rho_{1,4}(t) \\
0 & \rho_{2,2}(t) & 0 & 0 \\
0 & 0 & \rho_{3,3}(t) & 0 \\
\rho_{4,1}(t) & 0 & 0 & \rho_{4,4}(t)%
\end{array}%
\right),  \label{Eq:nv24}
\end{equation}
which is very convenient to evaluate entanglement measures such as the logarithmic negativity or concurrence \cite{Gauss1,Wootters98}. Though, for a qubit pair these two measures are equivalent, in this paper we focus on the concurrence ($C$). The initial two-qubit density operator corresponds to $\hat{\rho}(0)=|-,-\rangle \langle -,-|$, i.e. the qubit pair is in its ground state . We have solved analytically the master equation for the qubits in the stationary regimen and evaluated consequently the concurrence.

The stationary solution for the density matrix $\hat{\rho}^{ss}$ can be found analytically,
which for the density matrix given in Eq.~\eqref{Eq:nv24}, yields to $\rho_{2,2}^{ss}=\rho_{3,3}^{ss}$ and
\begin{eqnarray}
\nonumber C^{ss}&=&2Max\{ 0, |\rho_{1,4}^{ss}|-\rho_{2,2}^{ss}\}\\
\nonumber &=&\,{\rm Max}\{ 0\,,\,\frac{2\gamma {\rm tanh}[2s(\Delta)]-(\gamma^2-1){\rm sinh}^2[2s(\Delta)]}{(\gamma^2+1)+(\gamma^2-1){\rm cosh}[4s(\Delta)]} \}\\
\label{Eq:nv25}
\end{eqnarray}%
where $\gamma=(\gamma_1+\gamma_2)/2$ with $\gamma_j=\Gamma_j/\Gamma$ (from now on we shall simply denote $\Gamma=\Gamma_{1,2}$). Note that the steady-state two-qubit reduced density matrix, and consequently $C^{ss}$, does not depend separately on the individual dissipation rates $\gamma_j$ but only on their average value $\gamma$. Additionally, it is straightforward to verify that by putting $s_0=0$ in Eq.~\eqref{Eq:nv25}, i.e. non-squeezed microwave baths, each qubit is directly coupled to a local vacuum or zero temperature reservoir with no crossed arm couplings, producing a long-time diagonal density operator with $\rho^{ss}_{1,1}=\rho^{ss}_{2,2}=\rho^{ss}_{3,3}=\rho^{ss}_{1,4}=0$ and $\rho^{ss}_{4,4}=1$, corresponding to a vanishing qubit pair entanglement, $C^{ss}=0$.

\begin{figure}[t]
\begin{center}
\includegraphics[width=0.5\textwidth]{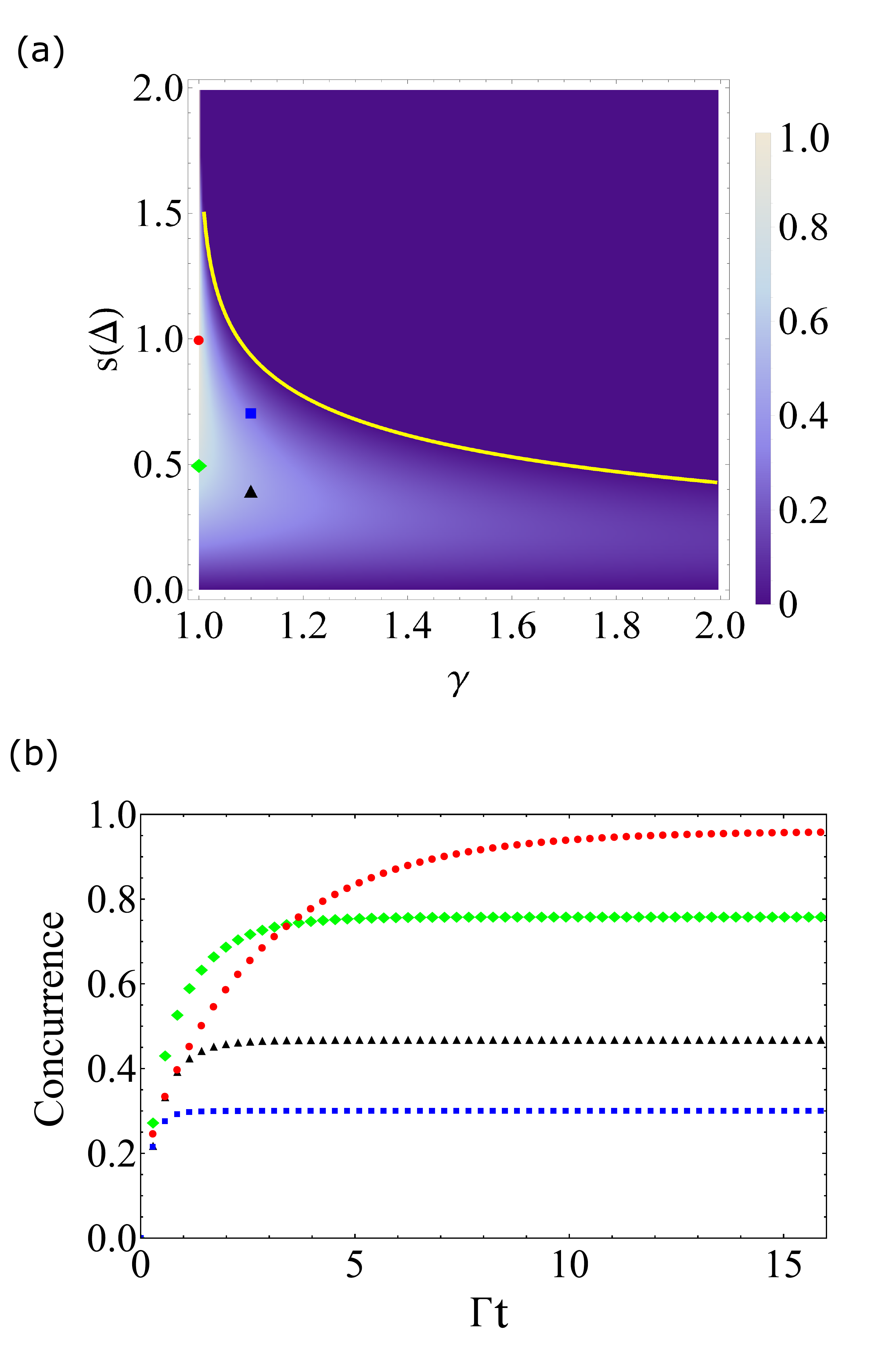}
\end{center}
\caption{Two-qubit concurrence $C$: (a) Steady-state $C^{ss}$ as a function of the microwave squeezing parameter $s(\Delta)$ and $\gamma=(\gamma_1+\gamma_2)/2$; the solid (yellow) line corresponds to Eq.~\eqref{Eq:lim1}. (b) Time dependent $C(t)$ for selected parameters as marked by similar symbols in (a) for two initially unentangled qubits.}
\label{fig:3}
\end{figure}

The result expressed by Eq.~\eqref{Eq:nv25} holds whenever $\gamma \geq 1$, otherwise the Lindblad master equation given by the set of Eqs.~\eqref{Eq:n1},~\eqref{Eq:nnn2} and ~\eqref{Eq:n2} does not possess steady-state solutions. The simplicity of this result allows us to obtain an analytical expression for the borderline in the parameter plane $(\gamma,s(\Delta))$
\begin{equation}
{\rm sinh}[4s(\Delta)]=\frac{4\gamma}{\gamma^{2}-1} ,\label{Eq:lim1}
\end{equation}
as shown by the yellow line in Fig.~\ref{fig:3}(a), separating regions of zero steady-state concurrence from regions of finite steady-state entanglement. As depicted in Fig.~\ref{fig:3}(a), if the microwave squeezing parameter $s(\Delta)$ increases the steady-state concurrence is also increased, but in order to reach this steady-state value a longer time is required.
For the special case $\gamma=1$, i.e. identical local average and non-local cross dissipation rates, we found $C^{ss}={\rm tanh}[2s(\Delta)]$ indicating that for a near resonance condition, $\Delta \approx 0$, and large microwave squeezing $r \gg 1$, the stationary concurrence gets saturated to its maximum value, $C^{ss}\rightarrow 1$, which corresponds to the qubit pair state approaching the pure Bell state $|\Psi^{ss}\rangle \approx \frac{1}{\sqrt{2}}\left ( |+,+\rangle-|-,-\rangle \right )$.

The two-qubit concurrence time evolution, $C(t)$, is depicted in Fig.~\ref{fig:3}(b) for some selected values of the squeezing parameter $s(\Delta)$ and local dissipation $\gamma$ terms, as marked by symbols in Fig.~\ref{fig:3}(a). Although the precise time evolution of $C(t)$ does depend on the individual values of $\gamma_j$, from now on we restrict ourselves to illustrate results only for the symmetric case, i.e. $\gamma_1=\gamma_2$. In all cases the concurrence starts growing linearly in time, i.e. $C(t)\sim t$, at short times. For the special line $\gamma=1$ in Fig.~\ref{fig:3}(a) the steady-state concurrence value, $C^{ss}$, requires longer times to be reached as the squeezing parameter $s(\Delta)$ gets larger.

\begin{figure}[t]
\begin{center}
\includegraphics[width=0.4\textwidth]{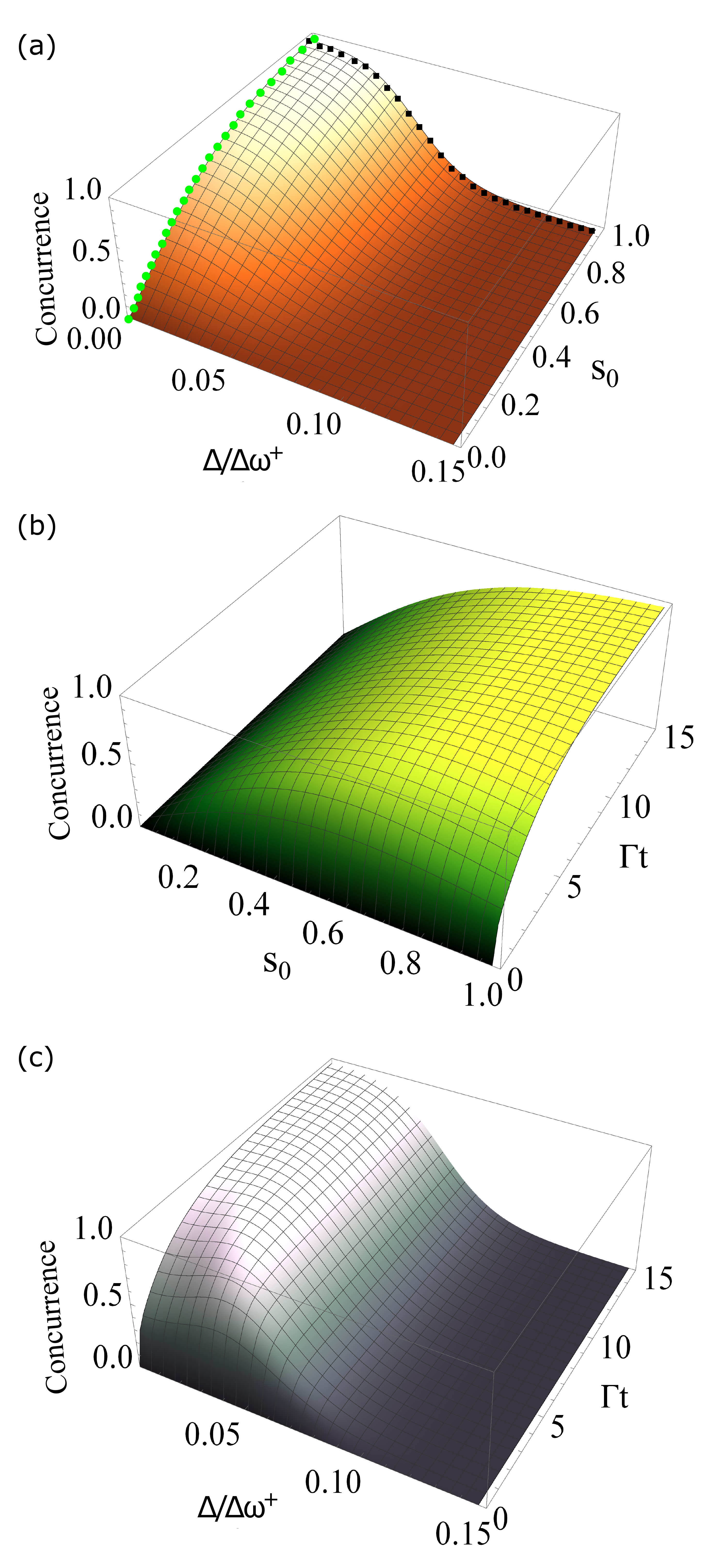}
\end{center}
\caption{Concurrence for two initially unentangled qubits for $\gamma=1$: (a) Steady-state $C^{ss}$ as a function of the resonance squeezing strength $s_0$ and the central microwave frequency-qubit detuning, $\Delta$. (b) Time dependent $C(t)$ at zero detuning, $\Delta=0$ (green points in (a)). (c) Time dependent $C(t)$ at fixed maximum squeezing $s_0=1$ (black points in (a)).}
\label{fig:4}
\end{figure}

Next, we discuss the two-qubit entanglement generation process from an unentangled initial two-qubit state, by considering separately the effects of the microwave resonance squeezing strength $s_0$ and the detuning between central microwave frequency and qubit splitting, $\Delta$. In Fig.~\ref{fig:4}(a), the steady-state concurrence $C^{ss}$ for $\gamma=1$ is plotted as a function of the resonance squeezing strength $s_0$ and the central microwave frequency-qubit detuning, $\Delta$. The time dependence of $C(t)$ is shown in Fig.~\ref{fig:4}(b) for the zero-detuning case, i.e. $\Delta=0$, as a function of the squeezing parameter $s_0$ (green points in Fig.~\ref{fig:4}(a)) while in Fig.~\ref{fig:4}(c) $C(t)$ is shown for the special microwave squeezing amount $s_0=1$ as a function of the detuning $\Delta$ (black points in Fig.~\ref{fig:4}(a)). In Fig.~\ref{fig:4}(c) it is also evident a two-time entanglement evolution: a fast entanglement generation at short times with $C(t)\sim t$ followed by a slower time-evolution towards the saturation value, a clear behavior especially seen near resonance.

\begin{figure}[t]
\begin{center}
\includegraphics[width=0.43\textwidth]{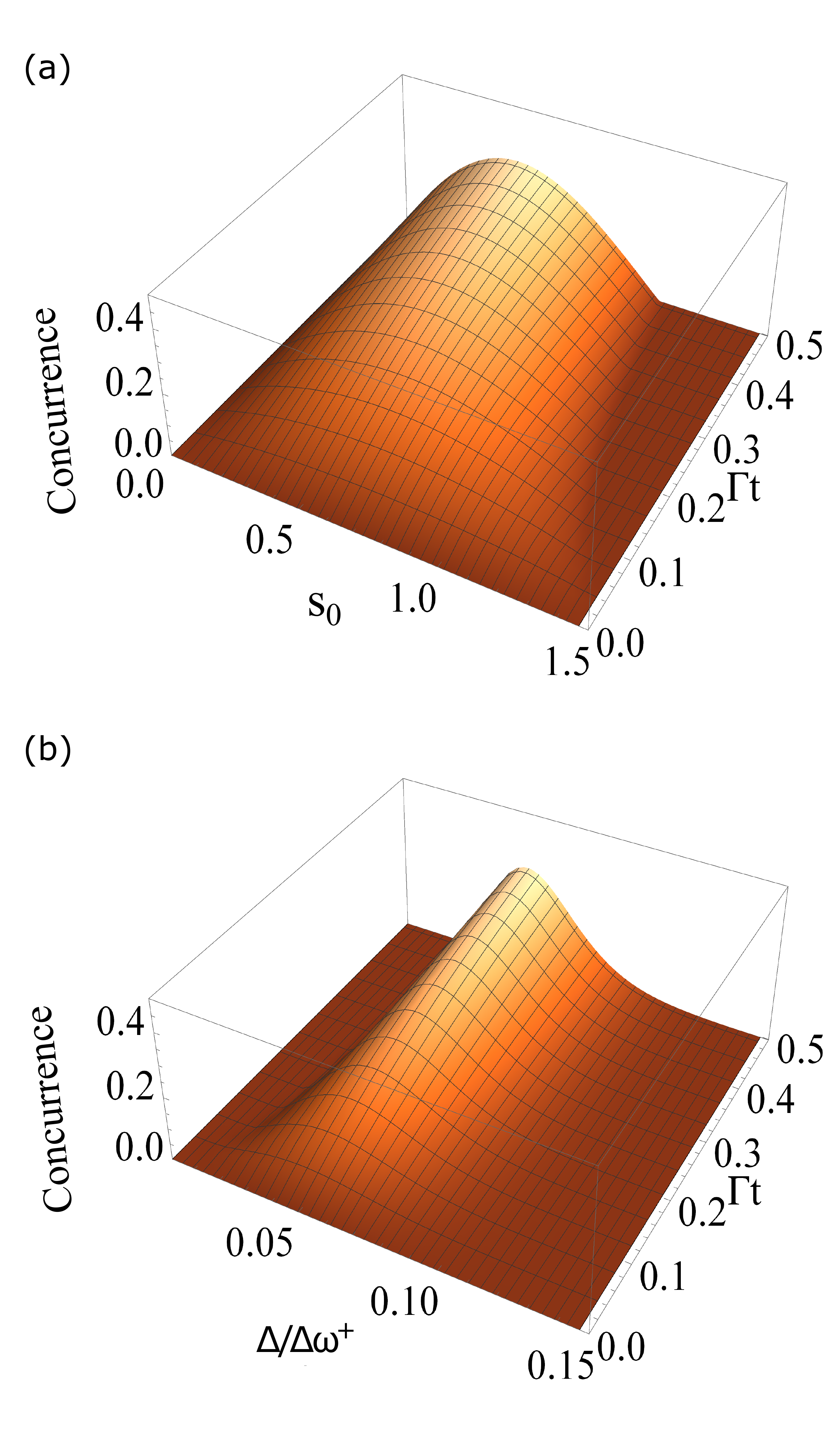}
\end{center}
\caption{Time dependent two qubit entanglement generation, $C(t)$, for initially unentangled qubits with symmetric local dissipation rates larger than the cross dissipation rate, $\gamma=1.1$: (a) Resonance, $\Delta=0$. (b) Fixed maximum squeezing parameter $s_0=1.5$.
}
\label{fig:5}
\end{figure}

In order to further explore the entanglement generation process we now consider some points in the parameter space (see Fig.~\ref{fig:3}(a)) outside the special $\gamma=1$ line. The two qubit time dependent concurrence $C(t)$ is depicted in Fig.~\ref{fig:5} for symmetric local dissipation rates larger than the non-local or cross dissipation rate, i.e. $\gamma=1.1$. Fig.~\ref{fig:5}(a) shows results of $C(t)$ at resonance, $\Delta=0$, as a function of the squeezing parameter $s_0$, where it is clear that $C(t)$ goes rapidly to its steady-state value for a microwave squeezing $s_0\approx 0.5$. However, at variance with the time evolution behavior for $\gamma=1$ (see Fig.~\ref{fig:4}(b)) for which the larger $s_0$ the larger the steady-state entanglement generation, now for $\gamma=1.1$ there is only a transient generation of $C(t)$ with a vanishing steady-state limit as previously shown in Fig.~\ref{fig:3}(a). An interesting behavior is uncovered by the plot in Fig.~\ref{fig:5}(b): by starting with a high microwave entanglement as corresponds to $s_0=1.5$ {\it and} the resonance condition $\Delta=0$ the two-qubit concurrence remains stuck to zero (both transient and steady-state values vanish, see Fig.~\ref{fig:3}(a); however, by detuning the qubit-microwave interaction, i.e. letting $\Delta > 0$, the effective squeezing parameter decreases $s(\Delta) < s_0$ allowing for the generation of two-qubit entanglement as can also been seen as a process where one starts at a point above the yellow line in Fig.~\ref{fig:3}(a) and by varying enough the detuning one crosses the yellow line to the zone of the parameter space where a steady-state entanglement is allowed. Thus, this behavior could mitigate the necessity of a special $\gamma$ value to generate a finite two qubit entanglement.

The crucial result of this section is that effectively it is possible to entangle distant qubits initially prepared in a separable state by using two entangled broadband microwave baths.

\section{BIPARTITE QUANTUM SYSTEM: TWO NON-INTERACTING QUBIT ENSEMBLES}
\label{sec:ensembles}

In the last section we have shown that a finite amount of entanglement of two initially separated qubits can be generated from two highly correlated microwave baths. However, this transfer process is strongly limited by the coupling strength between the qubits and the microwave radiation. In order to increase the matter-radiation coupling we propose to replace the system of two single qubits by two spin ensembles. In this way the matter-radiation coupling increases
\begin{figure}[t!]
\begin{center}
\includegraphics[width=0.48\textwidth]{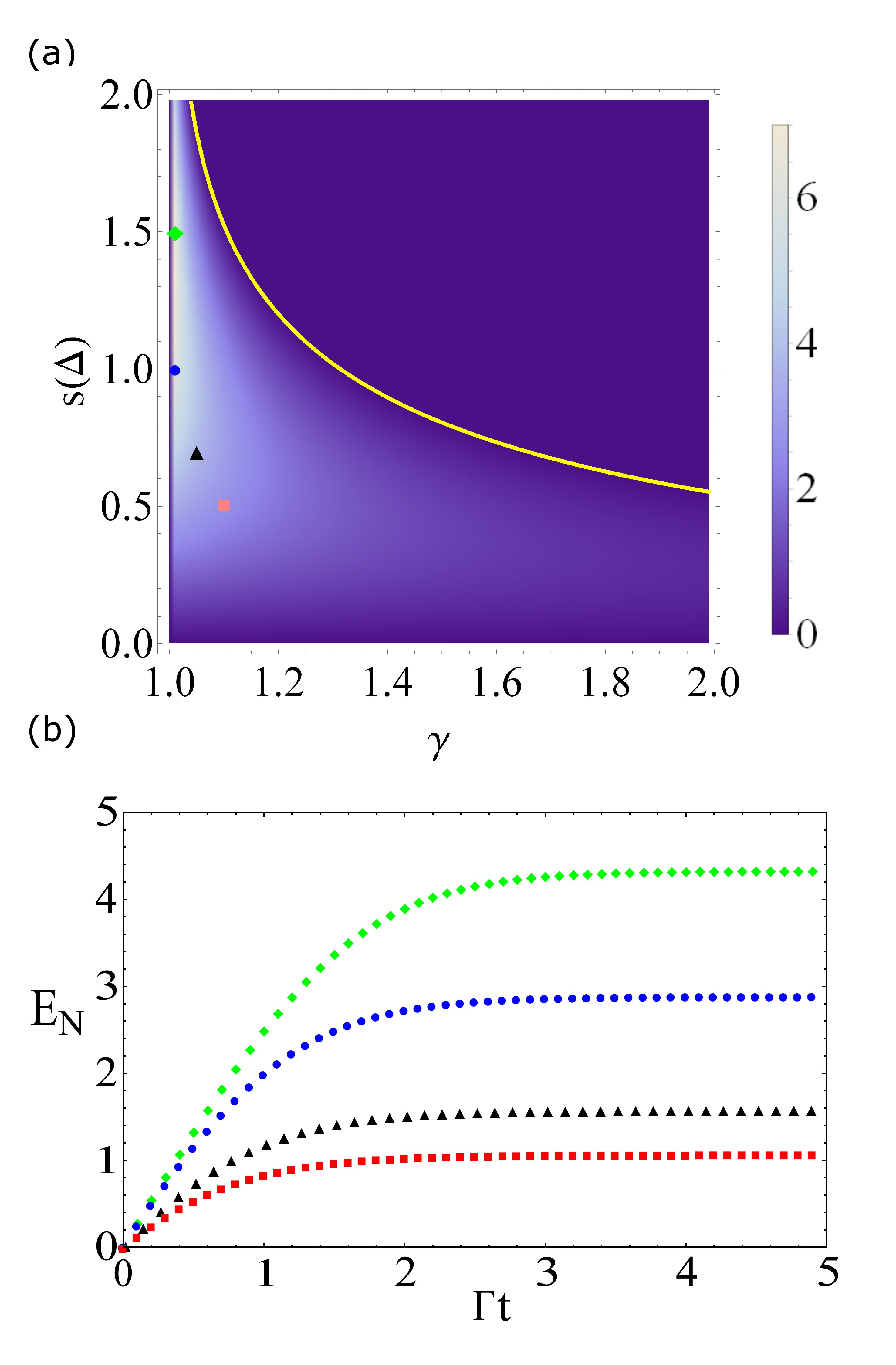}
\end{center}
\caption{Logarithmic negativity, $E_N$, for two spin ensembles: (a) Steady-state $E_N^{ss}$ as a function of the microwave squeezing parameter $s(\Delta)$ and $\gamma$; the solid (yellow) line corresponds to Eq.~\eqref{eq:EN_line}. (b) Time dependent $E_N(t)$ for selected parameters as marked by similar symbols in (a) for two initially unentangled spin ensembles.}
\label{fig:6}
\end{figure}
\begin{figure}[t!]
\begin{center}
\includegraphics[width=0.4\textwidth]{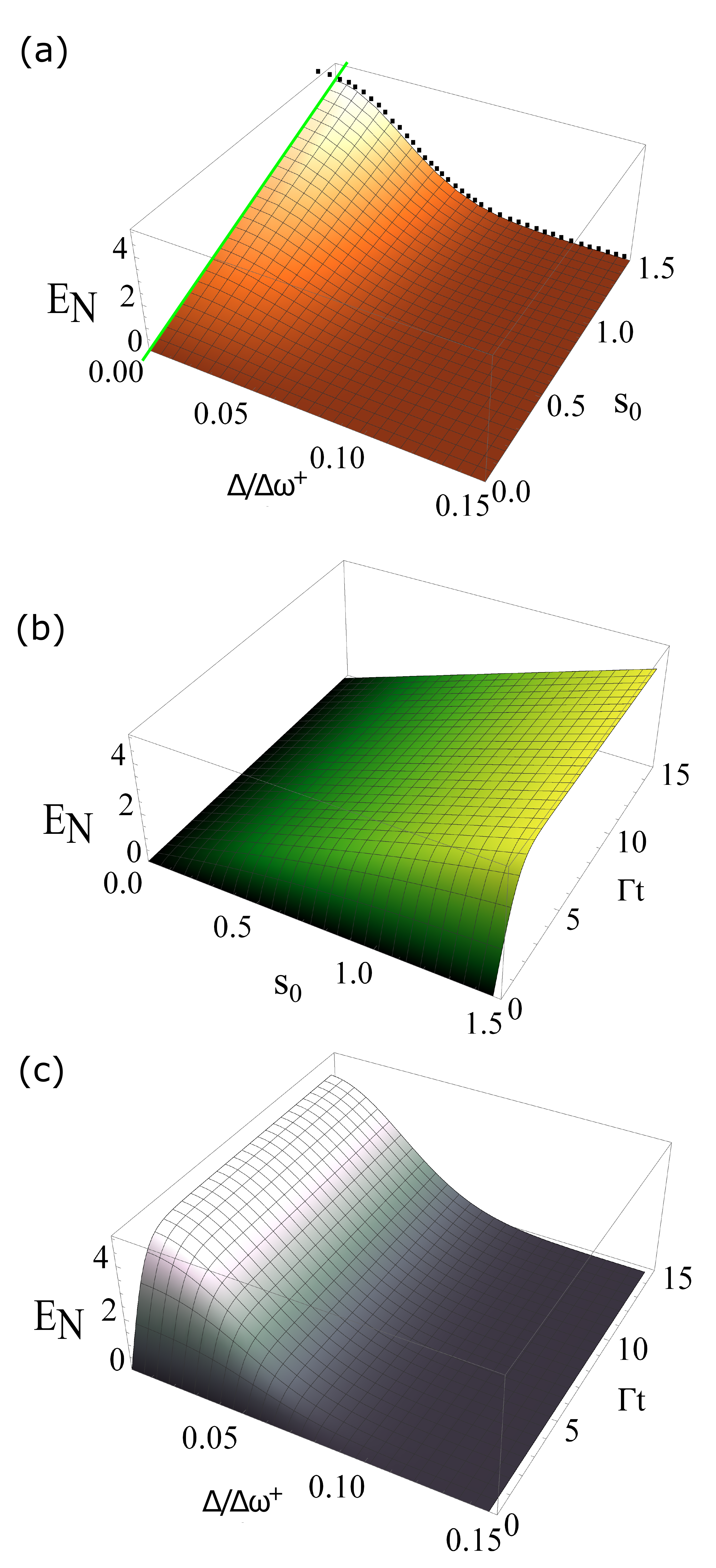}
\end{center}
\caption{Logarithmic negativity for two initially uncoupled qubit ensembles for $\gamma=1$: (a) Steady-state $E_N^{ss}$ as a function of the maximum squeezing strength $s_0$ and the central microwave frequency-qubit detuning, $\Delta$. (b) Time dependent $E_N(t)$ at resonance, $\Delta=0$ (green points in (a)). (c) Time dependent $E_N(t)$ at fixed maximum squeezing $s_0=1.5$ (black points in (a)).}
\label{fig:7}
\end{figure}
\begin{figure}[t!]
\begin{center}
\includegraphics[width=0.48\textwidth]{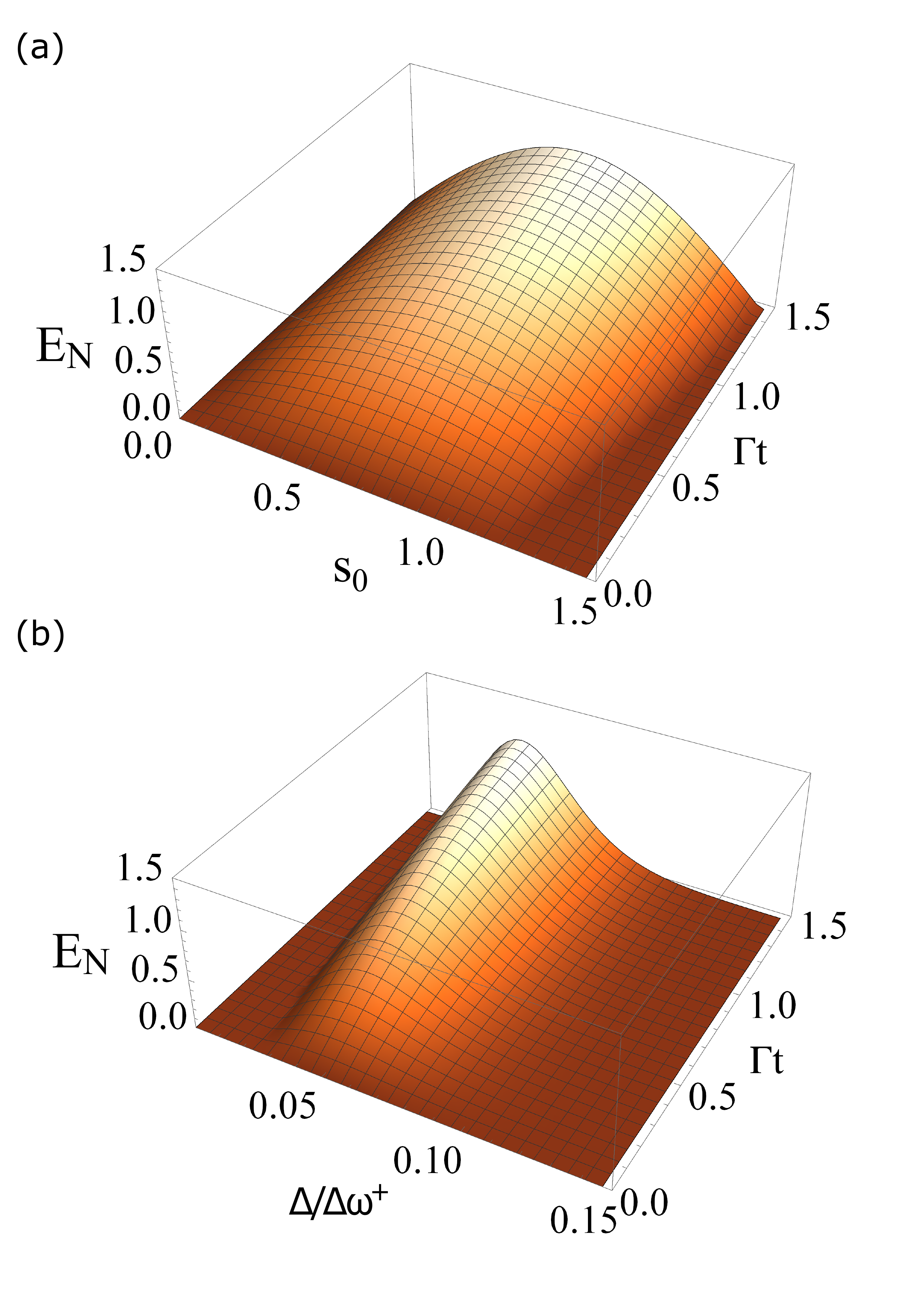}
\end{center}
\caption{Time dependent logarithmic negativity, $E_N(t)$, for initially unentangled qubit ensembles with symmetric local dissipation rates larger than the cross dissipation rate, $\gamma=1.1$: (a) Resonance, $\Delta=0$. (b) Fixed maximum squeezing parameter $s_0=1.5$.}
\label{fig:8}
\end{figure}
as $g\sim g_{e}\sqrt{N}$, where $g_{e}$ is the coupling of a single qubit with the microwave photons, yielding to an absolute increase of both local and non-local dissipation rates $\Gamma_j$ and $\Gamma$. An immediate positive consequence of this enhancement is to cut down the rise time of the entanglement generation up to reach the steady-state final value for two spin ensembles.

Assuming that each spin ensemble has a low polarization, i.e. they remain close to its global ground state, we can introduce collective bosonic operators associated to each qubit ensemble  $\hat{b}_{j}$,$\hat{b}^{\dagger}_{j}$, $j=1,2$~\cite{Collective_modes}. Thus, now we consider a central quantum system formed by two independent single-mode boson fields each of them coupled to a different reservoir of microwave radiation, but as before these microwave reservoirs still stay in a broadband squeezed multi-mode state. The master equation for the spin ensembles has the same structure as for single qubits, given by Eq.~\eqref{Eq:n1}, with the Lindblad terms as given in Eq.~\eqref{Eq:nnn2} and Eq.~\eqref{Eq:n2} but now with the replacements $\hat{q}^+_j \to \hat{b}^\dagger_j$ and $\hat{q}^-_j\to \hat{b}_j$ ($j=1,2$).

We study the dynamics of a subsystem composed of two initially non-interacting spin ensembles coupled to a broadband TLSMF from a JPA. We are interested in the time evolution of the degree of entanglement of the spin ensembles, having initially zero excitations i.e. $\hat{\rho}(0)=|0,0\rangle \langle 0,0|$, once they have interacted with the entangled microwaves. The state for the pair of spin ensembles is entirely specified by its covariance matrix, which is a real, symmetric and positive matrix ~\cite{Gauss1,Gauss2,Gauss3}
\begin{equation}
\hat{\sigma} (t)=\left(
\begin{array}{cccc}
\sigma _{xx} & \sigma _{xp_{x}} & \sigma _{xy} & \sigma _{xp_{y}} \\
\sigma _{xp_{x}} & \sigma _{p_{x}p_{x}} & \sigma _{yp_{x}} & \sigma
_{p_{x}p_{y}} \\
\sigma _{xy} & \sigma _{yp_{x}} & \sigma _{yy} & \sigma _{yp_{y}} \\
\sigma _{xp_{y}} & \sigma _{p_{x}p_{y}} & \sigma _{yp_{y}} & \sigma
_{p_{y}p_{y}}%
\end{array}%
\right) \label{Eq:nv98}
\end{equation}
The entries of the last matrix, $\sigma_{\alpha\beta}$, with $\alpha,\beta=x,y,p_{x},p_{y}$ are given by
\begin{equation}
\sigma_{\alpha\beta}=\frac{1}{2}\braket{\hat{\alpha}\hat{\beta}+\hat{\beta}\hat{\alpha}}
-\braket{\hat{\alpha}} \braket{\hat{\beta}},
\end{equation}
and they are computed from the canonical boson spin ensemble operators as
\begin{equation}
\hat{x}_{j}=\frac{(\hat{b}_{j}+\hat{b}_{j}^{\dag })}{\sqrt{2}},\ \ \ \ \ \ \hat{p}_{j}=\frac{%
(\hat{b}_{j}-\hat{b}_{j}^{\dag })}{i\sqrt{2}},
\end{equation}
with $(\hat{x}_{1},\hat{p}_{1})=(\hat{x},\hat{p}_{x})$ and $(\hat{x}_{2},\hat{p}_{2})=(\hat{y},\hat{p}_{y})$.

The entanglement of a two-mode Gaussian state is measured by the logarithmic negativity $E_{N}$, which has a closed expression given by
\begin{equation}
E_{N}=\max \left\{ 0,-\log _{2}2\tilde{\nu}_{-}\right\} \label{Eq:nv99}
\end{equation}%
Here, $\tilde{\nu}_{-}$ represents the smallest of the two symplectic eigenvalues of the partial transpose $\tilde{\sigma}$ of the two-mode covariance matrix $\hat{\sigma}$. Writing the matrix $\hat{\sigma}$ in terms of $2\times2$ blocks,
\begin{equation}
\sigma\equiv \left(
\begin{array}{cc}
\hat{A} & \hat{C} \\
\hat{C}^{T} & \hat{B}%
\end{array}%
\right),
\end{equation}
the logarithmic negativity in Eq.~\eqref{Eq:nv99} reads as
\begin{eqnarray}
E_{N}(t)&=&\max \left\{ 0,-\frac{1}{2}\log _{2}\left[ 4G(\sigma (t))\right]
\right\},
\end{eqnarray}

\begin{eqnarray}
G(\sigma) &=&\frac{1}{2}(\det A+\det B)-\det C- \nonumber\\
&&\left( \left[\frac{1}{2} \left( \det A+\det B\right) -\det C\right]
^{2}-\det \sigma\right) ^{1/2}.
\end{eqnarray}

As in the previous Section, for two qubit ensembles a well behaved steady-state solution of the Lindblad equation only exists for $\gamma \geq 1$.  Thus, in the stationary regime ($t\to\infty$) the only nonzero entries of the covariance matrix in Eq.~\eqref{Eq:nv98} are
\begin{eqnarray}
\sigma _{xx} &=&\sigma _{yy}=\frac{1}{2}\cosh [2s(\Delta)] = \sigma _{p_{x}p_{x}} = \sigma _{p_{y}p_{y}} \\
\sigma _{xy} &=&-\frac{1}{2\gamma}\sinh [2s(\Delta)] = -\sigma _{p_{x}p_{y}},  \notag \\
\end{eqnarray}
yielding to the following close expression for the steady-state logarithmic negativity:
\begin{eqnarray}
\nonumber E_{N}^{ss}=Max\{0,\log _{2}\left [\frac{\gamma}{\gamma \cosh[2s(\Delta )]-\sinh[2s(\Delta )]}\right ]\}\\
\label{eq:EN}
\end{eqnarray}
a result plotted in Fig.~\ref{fig:6}(a) where the borderline separating finite from zero $E_{N}^{ss}$ regions (solid yellow line) is now given by
\begin{equation}
{\rm sinh}[2s(\Delta)]=\frac{2\gamma}{\gamma^2-1} \label{eq:EN_line}
\end{equation}
An immediate comparison of this last result with that expressed by Eq.~\eqref{Eq:lim1} for the two-qubit case, leads us to conclude that the generation of steady-state entanglement between two qubit ensembles is allowed in a wider region of the parameter space $(\gamma,s(\Delta))$ than for a single qubit pair. In other words, there are parameter points for which two qubit steady-state entanglement never exists but, for the same parameter set, two qubit ensembles can indeed get entangled. For selected points marked in Fig.~\ref{fig:6}(a) the ensembles entanglement evolution $E_N(t)$ is plotted in Fig.~\ref{fig:6}(b) where it can be seen that for any parameter set, $E_N(t)\sim t$ with a single time constant.

The behavior of $E_N$, both steady-state and time dependent, as a function separately of the maximum microwave squeezing parameter $s_0$ and detuning $\Delta$, for $\gamma=1$, is shown in Fig.~\ref{fig:7}. Obviously, at resonance $E_N^{ss}$ grows boundless as a function of $s_0$ (see Fig.~\ref{fig:7}(a)), however with a fast degrading as the resonance condition is lost. For the time dependent behavior, we found a single time entanglement generation for any resonance detuning (see Figs.~\ref{fig:7}(b)-(c)).

Entanglement generation results for dissipation rates outside the special line $\gamma=1$ in Fig.~\ref{fig:6}(a) show a similar behavior as that reported in Section III for a qubit pair. Again notice that by increasing the detuning $\Delta$ one can cross the yellow line in Fig.~\ref{fig:6}(a), from above to below, allowing the qubit ensembles to become entangled at non-resonance conditions as shown in Figs.~\ref{fig:8}(a)-(b).

\section{THERMAL DECOHERENCE EFFECTS}
 \label{sec:decoherence}
The open quantum aspects of the results discussed so far have been limited to the TLSMF entangled reservoirs action upon the matter systems. However, the matter qubits in realistic solid-state setups are also exposed to other interactions with different degrees of freedom within the material or with additional external radiation fields which yield to a matter entanglement decreasing, though the rate of the entanglement generation by the TLSMF reservoirs themselves remain unaltered. Therefore, it is necessary to quantify the effects of realistic decoherence processes in the present systems of interest: NV centers, magnetic molecules and superconducting qubits. We shall concentrate on amplitude damping processes associated with thermal excitations as they constitute the main source of quantum correlation losses in condensed matter qubit systems ~\cite{Blais04}. However, if we focus on a specific solid-state realization a more detailed decoherence modeling might be required. Neighboring spins for NV centers ~\cite{Lange2010}, thermal fluctuations of dipolar interactions for magnetic molecules ~\cite{Dobrovitski20} and changes in the magnetic flux or external currents for superconducting qubits ~\cite{Yoshihara06} are some particular examples of decoherence processes in different condensed matter systems. If we include the thermal excitations the full master equation takes the form
 \begin{equation}
 \frac{d\hat{\rho}}{dt}=\hat{\mathcal{L}}_{MW}\hat{\rho}(t)+\hat{\mathcal{L}}_{D}\hat{\rho}(t) \label{Eq:nm1}.
 \end{equation}
 where $\hat{\mathcal{L}}_{MW}$ is given by Eqs.~\eqref{Eq:n1}-~\eqref{Eq:n2} and $\hat{\mathcal{L}}_{D}$ in Eq.~\eqref{Eq:nm1} represents the decoherence Lindbladian term associated with amplitude or thermal damping processes as given by:

 \begin{widetext}\begin{eqnarray}
 \hat{\mathcal{L}}_{D}\hat{\rho}(t)&=&\sum_{j=1,2}\Gamma_{j}^{th}\left \{ \left ( n_j+1 \right )\left [ 2\hat{q}_j^-\hat{\rho}(t)\hat{q}_j^+-\hat{q}_j^+\hat{q}_j^-\hat{\rho}(t)-\hat{\rho}(t)\hat{q}_j^+\hat{q}_j^- \right ] +
 n_j\left [2\hat{q}_j^+\hat{\rho}(t)\hat{q}_j^- -\hat{q}_j^-\hat{q}_j^+\hat{\rho}(t)-\hat{\rho}(t)\hat{q}_j^-\hat{q}_j^+ \right ] \right \}
 \label{Eq:decoh1}
 \end{eqnarray}\end{widetext}
 where $n_j=\left ( e^{\frac{\omega_j}{K_B T_j}}-1 \right )^{-1}$ denotes the Bose-Einstein occupation number for frequency $\omega_j$ of a thermal bath at temperature $T_j$ ($K_B$ is the Boltzmann constant) and $\Gamma_{j}^{th}$ represents a Weisskopf-Wigner effective thermal decay rate, for each $j=1,2$ matter subsystem. Notice that the decoherence/thermal Lindbladian $\hat{\mathcal{L}}_{D}$ in Eq.~\eqref{Eq:decoh1} has the same structure as the sum of local Lindbladians for the TLSMF case as given by Eq.~\eqref{Eq:nnn2}. Thus, we can conclude that the presence of incoherent thermal processes affecting the spin/qubit systems will leave unaffected the non-local entangling TLSMF Lindblad term in Eq.~\eqref{Eq:n2}. However, it is now expected a stronger competition between entangling and non-entangling terms.

 We shall concentrate upon discussing decoherence effects just for the steady-state entanglement behavior. For the two separate qubit case the border line separating the entangled from unentangled steady-state regions can now be written as ($\Gamma_{1}^{th}=\Gamma_{2}^{th}=\Gamma^{th}$ and $n_1=n_2=n$):
 \begin{eqnarray}
 \sum\limits_{i=0}^6 p_i(\gamma,\gamma_{th},n) x^i=0
 \label{Eq:decoh2}
 \end{eqnarray}
 with $x=exp\{ 2s(\Delta) \}$ and $\gamma_{th}=\Gamma^{th}/\Gamma_{1,2}$. The explicit forms of functions $p_i(\gamma,\gamma_{th},n)$ are given in Appendix B where it is easily confirmed that in case of $\gamma_{th}=0$ it follows that $p_1=p_3=p_5=0$ retrieving the expression previously given
 in Eq.~\eqref{Eq:lim1}. For qubit ensembles in presence of thermal effects the region in the $s(\Delta)-\gamma$ parameter plane of finite steady-state logarithmic negativity $E_N$, is now given by the following inequalities:
 \begin{eqnarray}
 p_-(\gamma,\gamma_{th},n) \leq exp\{ s(\Delta) \}\leq p_+(\gamma,\gamma_{th},n)
 \label{Eq:decoh3}
 \end{eqnarray}
 with
 \begin{eqnarray}
 \nonumber p_{\pm}(\gamma,\gamma_{th},n)=\sqrt{\frac{\gamma -2n\gamma _{th}\pm \sqrt{1+4n\gamma
 _{th}(n\gamma _{th}-\gamma )}}{\gamma -1}}\\
 \label{Eq:decoh33}
 \end{eqnarray}
 Notice that when $n\gamma _{th}\rightarrow 0$ the border line separating entangled from unentangled zones is given by
 \begin{eqnarray}
 exp\{ s(\Delta) \}=\sqrt{\frac{\gamma +1}{\gamma -1}}
 \label{Eq:decoh3p}
 \end{eqnarray}
 which yields directly to Eq.~\eqref{eq:EN_line}.

\begin{figure}[t!]
 \begin{center}
 \includegraphics[width=0.5\textwidth]{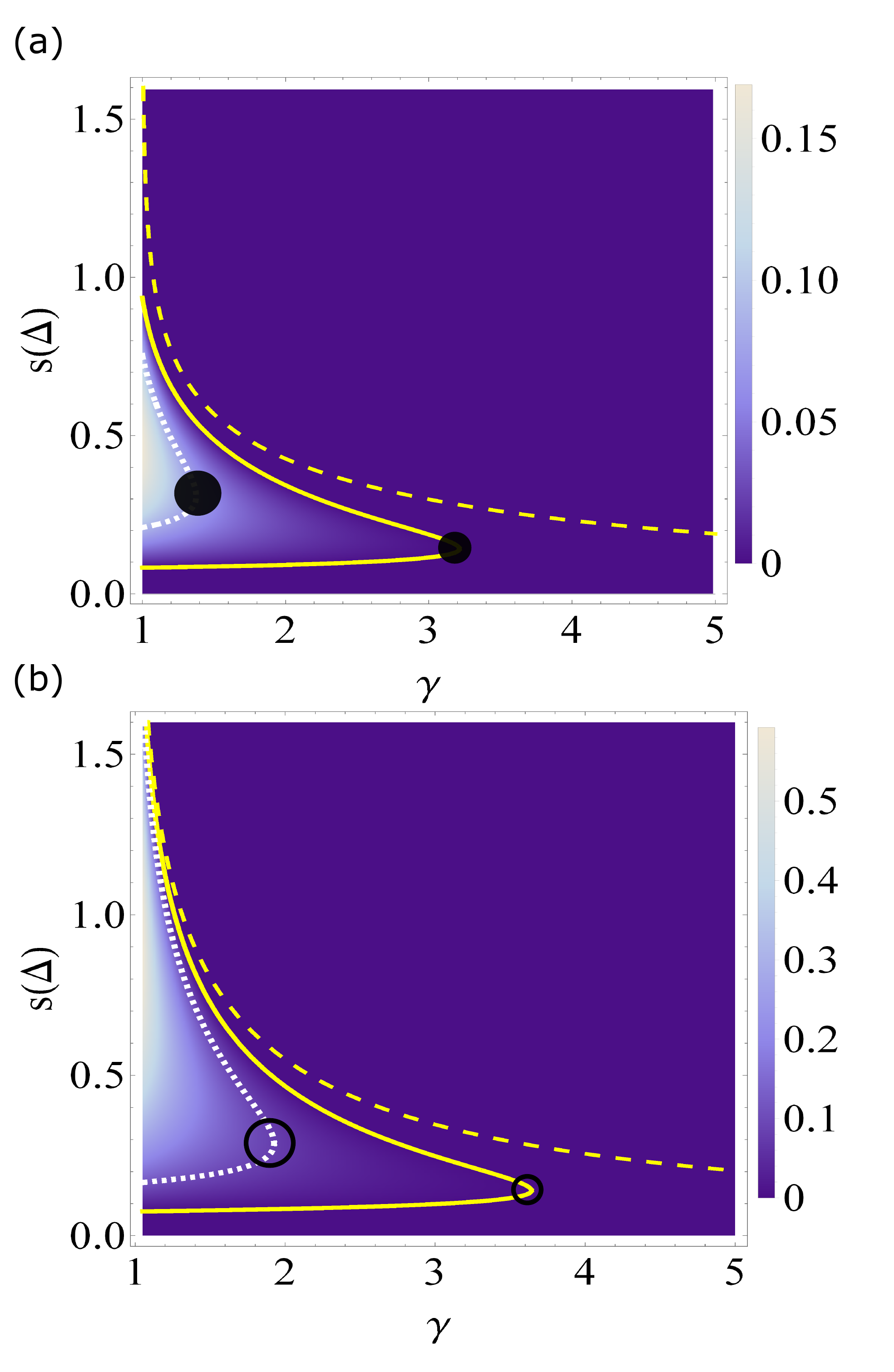}
 \end{center}
 \caption{Steady-state entanglement in the TLSMF parameter space $\{ s(\Delta), \gamma\}$: (a) two-qubit concurrence $C^{ss}$ and (b) two multi-spin logarithmic negativity, $E_N^{ss}$. For both frames (a)-(b) lines represent the border separating entangled from unentangled states: dashed (yellow) lines denote pure TLSMF entanglement transfer, i.e. $\gamma_{th}=0.0$, solid (yellow) lines correspond to thermal coupling $\gamma_{th}=0.7$ and mean thermal excitation number $n=0.1$ while dotted (white) lines denote a higher temperature case with $\gamma_{th}=0.7$ and $n=0.16$. Circles in the figures, with coordinates $\{ s^{*}(\Delta), \gamma^{*} \}$, represent the maximum $\gamma$ values, and corresponding squeezing $s(\Delta)$, to reach entangled states (see also Fig.~\ref{fig:12}).}
 \label{fig:10}
 \end{figure}

\begin{figure}[t!]
 \begin{center}
 \includegraphics[width=0.5\textwidth]{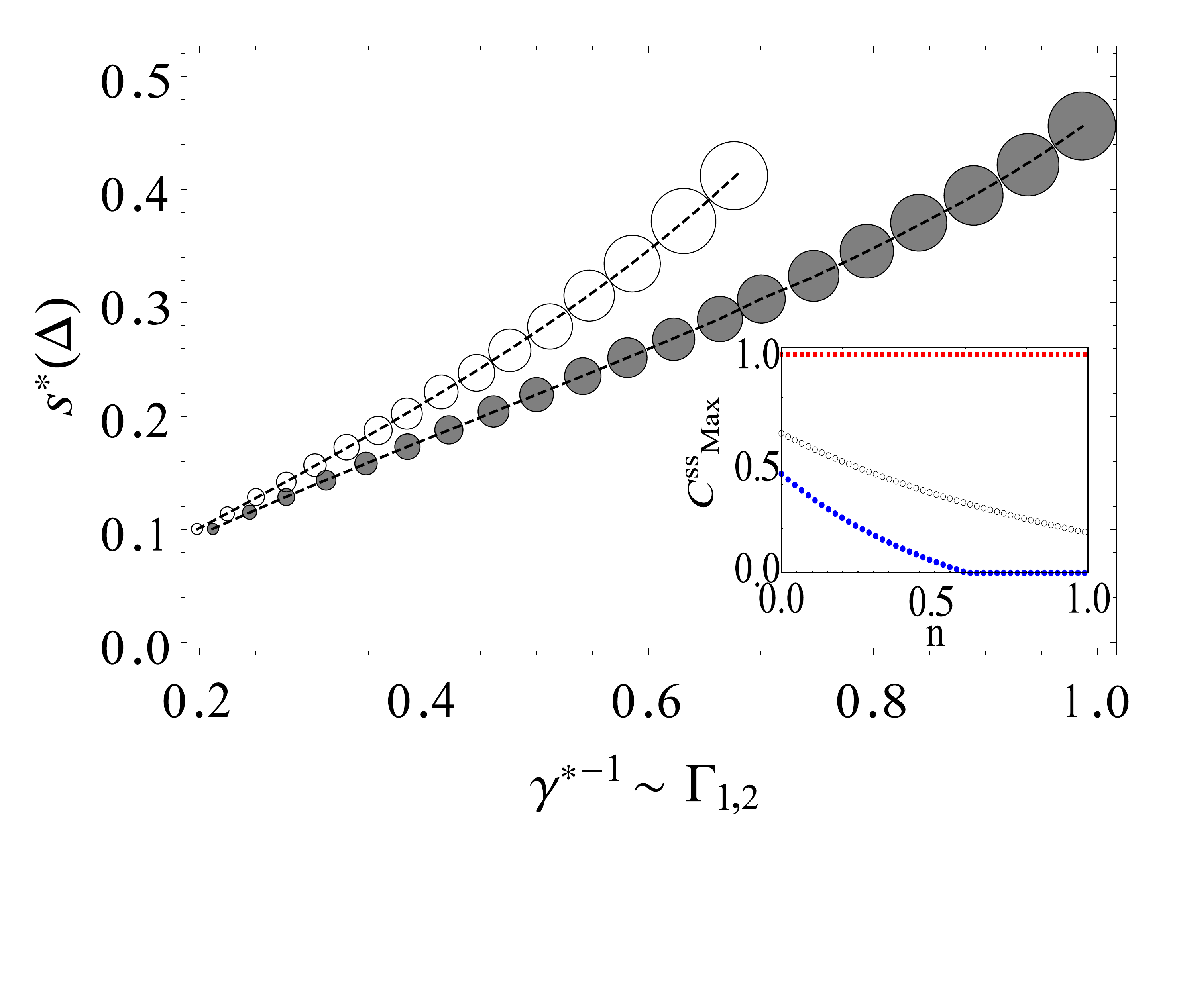}
 \end{center}
 \caption{Extreme values $s^{*}(\Delta)$ and $\gamma^{*-1} \sim \Gamma_{1,2}$ (see also Fig.~\ref{fig:10}) to obtain stationary entangled states for qubit pairs (filled circles) and qubit ensembles (empty circles) for a thermal coupling $\gamma_{th}=0.7$ as the temperature is varied producing mean excitation thermal numbers ranging from $n=0.05$ to $n=0.19$. The area of each circle is proportional to $n$. Inset: Maxima of steady-state concurrence for a qubit pair system as a function of $n$ (or temperature): red squares represent no thermal bath $\gamma_{th}=0$, empty black circles $\gamma_{th}=0.1$ and filled blue circles $\gamma_{th}=0.3$
 }
 \label{fig:12}
 \end{figure}

Thermal decoherence effects on the TLSMF mediated entanglement generation for both cases, a qubit pair and multi-qubit ensembles, are depicted in Fig.~\ref{fig:10}-(a,b) for $\gamma_{th}=0.7$ and $n=0.1$. First, by comparing the color scales in Fig.~\ref{fig:10}-(a,b) with the corresponding scales in Fig.~\ref{fig:3} and Fig.~\ref{fig:6}, we find that the maximum entanglement amount is reduced roughly one order of magnitude with respect to the pure TLSMF case. Second, a reduction of the entanglement zone when thermal effects are incorporated is also observed in both cases (the solid yellow lines correspond to the border line separating the steady-state entangled-unentangled regions in the $s(\Delta)-\gamma$ parameter plane). For the sake of comparison we also display with dashed yellow lines the $\gamma_{th}=0$ results as previously plotted in Fig.~\ref{fig:3}-(a) and Fig.~\ref{fig:6}-(a). For additional information we also plot, as white dotted lines, in Fig.~\ref{fig:10}-(a,b) the borders of the steady-state finite entanglement regions for a higher temperature situation, i.e. $\gamma_{th}=0.7$ and $n=0.16$. Notice that for low microwave squeezing values, small $s(\Delta)$, the steady-state entanglement disappears quickly as the crossed squeezing term $\Gamma_{1,2}$ in Eq.~\eqref{Eq:nv222} decreases or equivalently when $\gamma$ values increase. Furthermore, it is evident that thermal decoherence effects impose, as a requirement for generating steady-state entanglement, a minimum amount of microwave squeezing $s(\Delta)$. Notice also that a maximum value of $\gamma$ (or equivalently a minimum value of cross TLSMF rate $\Gamma_{1,2}$) as determined by each temperature, is required to achieve steady-state entanglement. These extreme values $({s^{*}(\Delta),\gamma^{*}})$ are represented in Fig.~\ref{fig:10} with filled circles for qubit pairs and empty circles for qubit ensembles. The dependence of these extreme values for a range of temperatures is depicted Fig.~\ref{fig:12}, where the area of the circles is proportional to the average number of excitations $n$. Evidently the larger the temperature a higher minimum value of cross TLSMF $\Gamma_{1,2}$ is required. The maximum value of concurrence for a qubit pair as a function of temperature ($n$) is shown in the inset of Fig.~\ref{fig:12}. Since the logarithmic negativity for the multi-qubit systems is not bounded a similar graph can not be built for different temperatures. By including other sources of dissipation the long time entanglement is further degraded, nevertheless the qualitative behavior described previously is still observed. The optimal values of $\Gamma_{1,2}$ and $s(\Delta)$ can be controlled to reach a significant amount of entanglement.

All the main results described for the $\gamma_{th}=0$ situation survive well if decoherence towards thermal environments is on the same order of magnitude as the cross TLSMF $\Gamma_{1,2}$ value and low enough temperatures. According with these results we conclude that thermal decoherence effects limit both the maximum
 amount of entanglement for the two cases considered as wells as the region in the $s(\Delta)-\gamma$ parameter space where a stationary entanglement can be reached. However, the existence of finite steady-state entanglement generated by TLSMF can survive the attack of thermal effects. Therefore the entanglement transfer from entangled microwaves to solid state qubits pairs or multi-qubit ensembles is a reliable and robust process even in the presence of noisy environments.

\section{EXPERIMENTAL IMPLEMENTATION}
\label{sec:implementation}
We have explained two theoretical protocols to transfer quantum correlations from a squeezed bath to two initially uncoupled systems, either qubits or ensembles that behave like effective harmonic oscillators. We will now discuss how these ideas can be adapted to a setup where the entangling bath is squeezed microwave radiation generated and propagating through a superconducting device.

JPA~\cite{Yurke87} are superconducting devices that, by combining an external driving with some incoming radiation, can produce huge signal gains with the addition of very minimal, quantum limited noise. At the same time that they amplify the radiation, these devices are also capable of producing very large amounts of squeezing, either on some income field or in pure vacuum. From early implementations with about 50\% noise reduction~\cite{Yurke91,Yurke88}, state-of-the-art implementations now reach values of 10dB squeezing in an input vacuum state~\cite{CastellanosBeltran08}, figures that improve every year. When operated in the frequency downconversion regime, JPA generates pairs of correlated photons in a two-mode squeezed state such as the one used in this work~\cite{Eichler11}. Alternatively to ordinary
JPA's we also find other Josephson devices in the literature which are specifically tuned for two-mode squeezing generation and which hold a greater potential for large squeezing values~\cite{Eichler14,Flurin12}, already facilitating values of 12dB two-mode squeezing~\cite{Eichler14}.

Let us first discuss the situation in which the TLSMF couples to an ensemble of NV-centers. The advantage of using spin ensembles composed by NV centers are the similar energy with the squeezed microwaves generated in quantum circuits. Each NV center has a $S=1$ ground state, with zero-field splitting $\Delta =2\pi \times 2.87\ GHz$ between the $m_{s}=0$ and $m_{s}=\pm 1$ states. By the application of an external magnetic field, one can isolate two spin transitions of this triplet due to the fact that the zero-field spin splitting $\Delta $ sets a preferred axis of quantization to be along the axis of the nitrogen-vacancy bond and model the NV like two qubits. For the coupling between the NV spin ensemble and the microwave field we have taken experimental reported parameters where for  $N=3\times 10^{7}$ colour centers $g\approx 2\pi \times 35MHz$~\cite{Zhu}. In order to quantify the entanglement between the ensembles once have interacted with the correlated baths we calculated the logarithmic negativity, in Sect.\ \ref{sec:ensembles} and we show the optimal parameter of squeezing of the microwaves $s_0$ required. In Fig.~\ref{fig:6} to Fig.~\ref{fig:8} we can note that appreciable entanglement is obtained for $s_0\leq 0.5$ in both regimes: time depending and stationary, this value corresponds to a gain $G_{E}=cosh^{2}[s_0] = 1.27$ dB, therefore the required squeezing for the microwaves to obtain maximum entangled values is in the range of the reported experimental values. The other essential parameter in that process is the $\Gamma$, where $\Gamma\sim g^2$. For the case of a spin ensemble $g\approx 2\pi \times 35$ MHz, and the results show that with this parameter we can obtain significant values of entanglement between the two spin ensembles. In the stationary regime the $E^{ss}_{N}$ function reaches the value $5$ for $s_0=1.5$.
The other experimental setup that we propose consist of two single qubits which are spatially separated and coupled to different transmission line modes. As we have seen above, a good qubit-radiation coupling is essential for a successful transfer of correlations, which may condition the implementation. If those qubits are NV-centers, the typical coupling to the microwaves will be rather small, about $100$ Hz for bare line, or slightly larger, $\sim0.1 $MHz, for more sophisticated coupling mechanisms~\cite{Twamley,Cheng}, but always on the border line and dominated by other dephasing or dissipation mechanisms. One interesting alternative is to rely on molecular magnets: still in the range of microwaves, these macromolecules host ions with large magnetic moments and they can be placed cleverly for enhanced coupling with the radiation. For example for molecules of $Fe_{8}$ the relation $g/\omega$ is three orders of magnitude greater than for NV centers~\cite{Juanjo}. Finally, the simplest route would be to use our protocol to entangle ordinary superconducting qubits. 
In QED circuit experiments it has already been demonstrated strong coupling of microwave photons confined in a transmission line cavity with single superconducting qubits, 
 with coupling strengths between matter and radiation reaching values of up to 
 $g=105 MHz$ ~\cite{Majer2007}, while thermal decay effects in such low-temperature setups are so weak that effects in such low-temperature setups are so weak that our results shown in 
 Fig.~\ref{fig:10} and  Fig.~\ref{fig:12} are indeed relevant.

\section{CONCLUSIONS}
\label{sec:summary}

Summing up, in this work we have proposed a hybrid system in order to study the dynamics of quantum correlations transferred to initially uncoupled single or spin ensembles from a squeezed microwave field generated by JPA devices combining recent advances on parametric amplifiers with NV centers and other solid-state spin systems which share similar energy scales to that of controlled microwave radiation fields. For the case of spin ensembles we evidence a notable value for the entanglement without requiring high values for the squeezing parameter of microwaves $r$, this facilitate the possible experimental implementation. And even more interesting is the possibility of get entanglement even in the stationary regimen. The experimental values for coupling between spin ensembles and microwaves show that this proposal would serve for obtain highly entangled states. In the case of the qubits is of great interest control single quantum particles and reach entanglement among them. The results show that if we combine the squeezed microwaves with single-molecule magnets or superconducting qubits is possible to reach non negligible spin entanglement values. We have established the robustness of TLSMF entanglement transfer processes under thermal dissipative conditions.
\acknowledgements

A.V.G., F.J.R. and L.Q. acknowledge financial support from Facultad de Ciencias at UniAndes-2015 project "Transfer of correlations from non-classically correlated reservoirs to solid state systems" and project "Quantum control of non-equilibrium hybrid systems-Part II", UniAndes-2015. J.J.G.R. acknowledges support from Spanish Mineco project FIS2012-33022, from EU FP7 project PROMISCE, from CAM Research Network QUITEMAD+.

\appendix
\section{Broadband squeezing transformations}
To further proceed, it is important to
realize that reservoir boson operators are transformed by the action of the squeezing operator $\hat{S}$ (see Eq.~\eqref{Eq:eq8}) as:
\begin{eqnarray}
\hat{S}^{\dag }\hat{a}_{n,j}\hat{S} &=&\sum_m\left [ \mathrm{A}^{(1)}_{n,m}\hat{a}
_{m,j}+\mathrm{A}^{(2)}_{n,m}\hat{a}_{m,\bar{j}}^{\dag}\right ]  \notag \\
\hat{S}^{\dag }\hat{a}_{n,\bar{j}}^{\dag }\hat{S} &=&\sum_m\left [ \mathrm{A}^{(1)}_{n,m}\hat{a}_{m,\bar{j}}^{\dag}+\mathrm{A}^{(2)}_{n,m}\hat{a}_{m,j}\right ]
\label{Eq:eq14}
\end{eqnarray}
$j,\bar{j}=1,2$ with $j+\bar{j}=3$. In Eq.(\ref{Eq:eq14}) the expressions for $\mathrm{A}^{(j)}_{n,m}$ are given by:
\begin{widetext}
\begin{eqnarray}
\mathrm{A}^{(1)}_{n,m}&=&\delta_{n,m}+\frac{1}{2!}\sum_{p}r_{n,p}r_{m,p}+
\frac{1}{4!}\sum_{p,q,r}r_{n,p}r_{q,p}r_{q,r}r_{m,r}+\cdot\cdot\cdot  \notag \\
\mathrm{A}^{(2)}_{n,m}&=&r_{n,m}+\frac{1}{3!}\sum_{p,q}r_{n,p}r_{q,p}r_{q,m}+
\frac{1}{5!}\sum_{p,q,r,s}r_{n,p}r_{q,p}r_{q,r}r_{s,r}R_{s,m}+\cdot\cdot\cdot
\label{Eq:eq114}
\end{eqnarray}
\end{widetext}
On inserting Eqs.(\ref{Eq:eq14}) into Eq.(\ref{Eq:eq13}) it follows that the only different from zero
partial traces over the reservoir degrees of freedom correspond to:
\begin{eqnarray}
\langle \hat{a}_{n,1}^{\dag }\hat{a}_{m,1}\rangle  &=&\langle \hat{a}_{n,2}^{\dag }\hat{a}_{m,2}\rangle=\sum_{p}\mathrm{A}^{(2)}_{n,p}\mathrm{A}^{(2)}_{m,p}  \notag \\
\langle \hat{a}_{n,1}^{\dag }\hat{a}_{m,2}^{\dag }\rangle  &=&\langle \hat{a}
_{n,1}\hat{a}_{m,2}\rangle =\sum_{p}\mathrm{A}^{(1)}_{n,p}\mathrm{A}^{(2)}_{m,p}  \notag \\
\langle \hat{a}_{n,1}\hat{a}_{m,1}^{\dag }\rangle  &=&\langle \hat{a}_{n,2}
\hat{a}_{m,2}^{\dag }\rangle =\sum_{p}\mathrm{A}^{(1)}_{n,p}\mathrm{A}^{(1)}_{m,p}  \label{Eq:eq15}
\end{eqnarray}

On inserting Eq.~\eqref{Eq:eq116}, valid for a perfect correlated two-bath system, into Eq.~\eqref{Eq:eq114} we arrive to the simple expressions
\begin{eqnarray}
\nonumber \mathrm{A}^{(1)}_{n,m}&=&\delta_{n,m}\mathrm{cosh}(r_{n})\\
\mathrm{A}^{(2)}_{n,m}&=&\delta_{n,m}\mathrm{sinh}(r_{n}) \label{Eq:eq117}
\end{eqnarray}
On substituting Eqs.~\eqref{Eq:eq117} into Eqs.~\eqref{Eq:eq14} we obtain Eqs.~\eqref{Eq:eq118}.

\section{Competition between squeezing and thermal effects}

Here we proceed to give the explicit form for the steady-state two-qubit concurrence $C^{ss}$. It is given by:
\begin{eqnarray}
C^{ss}=\frac{1}{4}\frac{\sum\limits_{i=0}^6 p_i(\gamma,\gamma_{th},n) x^i}{\sum\limits_{i=0}^6 q_i(\gamma,\gamma_{th},n) x^i}
\label{Eq:ab0}
\end{eqnarray}
with $x=exp\{ 2s(\Delta) \}$ and the explicit forms of functions $p_i(\gamma,\gamma_{th},n)$ and $q_i(\gamma,\gamma_{th},n)$ are:
\begin{widetext}\begin{eqnarray}
\nonumber p_0(\gamma,\gamma_{th},n)&=&p_6(\gamma,\gamma_{th},n)=-\gamma(\gamma^2-1)\\
\nonumber p_1(\gamma,\gamma_{th},n)&=&p_5(\gamma,\gamma_{th},n)=2\gamma_{th}(1+2 n)(1-3\gamma^2)\\
\nonumber p_2(\gamma,\gamma_{th},n)&=&-\gamma-8\gamma^2+\gamma^3-16\gamma\gamma_{th}+8\gamma^2\gamma_{th}-8\gamma_{th}^{2}-8\gamma\gamma_{th}^{2}-48 \gamma\gamma_{th}^{2}n(n+1)\\
\nonumber p_4(\gamma,\gamma_{th},n)&=&-\gamma+8\gamma^2+\gamma^3+16\gamma\gamma_{th}+8\gamma^2\gamma_{th}+8\gamma_{th}^{2}-8\gamma\gamma_{th}^{2}-48 \gamma\gamma_{th}^{2}n(n+1)\\
p_3(\gamma,\gamma_{th},n)&=&-4\gamma_{th}\left [ 1+\gamma^2-4\gamma\gamma_{th}+2(1+\gamma^2)n-8\gamma_{th} n\left (\gamma_{th}-\gamma+3\gamma_{th}n+2\gamma_{th}n^2 \right ) \right ]
\label{Eq:ab1}
\end{eqnarray}\end{widetext}
and
\begin{widetext}\begin{eqnarray}
\nonumber q_0(\gamma,\gamma_{th},n)&=&q_6(\gamma,\gamma_{th},n)=\gamma(\gamma^2-1)\\
\nonumber q_1(\gamma,\gamma_{th},n)&=&q_5(\gamma,\gamma_{th},n)=\gamma\left [ 1+3\gamma^2+12\gamma_{th}^{2}\left ( 1+2n \right )^2 \right ]\\
\nonumber q_2(\gamma,\gamma_{th},n)&=&q_4(\gamma,\gamma_{th},n)=-\gamma-8\gamma^2+\gamma^3-16\gamma\gamma_{th}+8\gamma^2\gamma_{th}-8\gamma_{th}^{2}-8\gamma\gamma_{th}^{2}-48 \gamma\gamma_{th}^{2}n(n+1)\\
q_3(\gamma,\gamma_{th},n)&=&16\gamma_{th}\left [ (1+2n)(1+3\gamma^2)+2\gamma_{th}^{2}\left (1+6n+12n^2+8n^3 \right ) \right ]
\label{Eq:ab2}
\end{eqnarray}\end{widetext}

\bibliography{mybib}	

\end{document}